\documentclass[12pt,a4paper,thmsb]{article}
\usepackage{eurosym}
\usepackage{amsmath}
\usepackage{amssymb}
\usepackage{amsfonts}
\usepackage{color}
\newtheorem{theorem}{Theorem}

\newtheorem{example}{Example}
\newenvironment{proof}[1][Proof]{\noindent\textbf{#1.} }{\ \rule{0.5em}{0.5em}}

\long\def\comment#1{}

\begin{document}
\renewcommand{\thefootnote}{\fnsymbol{footnote}}
\title{Productivity and quality-adjusted life years: QALYs, PALYs and beyond\footnotemark[1]}

\author{Kristian S. Hansen\footnotemark[2] 
\and Juan D. Moreno-Ternero\footnotemark[3]
\and Lars P. \O sterdal\footnotemark[4]}
\footnotetext[1]{
We thank Owen O'Donnell and two anonymous referees for helpful comments and suggestions. We also thank Ramses Abul Naga, Matt Adler, Harald Hannerz, Anita Marie Glenny, Alec Morton, as well as participants at the Roskilde Conference on Working Environment Economics, the Paris SAET Meeting, and the seminars at the Universidad de M\'{a}laga and the University of Hawai'i-Manoa. 
Financial support from the National Research Centre for the Working Environment (NFA), Copenhagen, Denmark, and the Spanish Agencia Estatal de Investigaci\'{o}n (AEI) through grant PID2020-115011GB-I00, funded by MCIN/AEI/10.13039/501100011033, is gratefully acknowledged.}
\footnotetext[2]{National Research Centre for the Working Environment (NFA), Copenhagen, Denmark. e-mail: krh@nfa.dk
}
\footnotetext[3]{Department of Economics, Universidad Pablo de Olavide. e-mail: jdmoreno@upo.es}
\footnotetext[4]{Department of Economics, Copenhagen Business School. e-mail: lpo.eco@cbs.dk  }


\date{
\today}

\maketitle

\renewcommand{\thefootnote}{\arabic{footnote}}

\begin{abstract}
We develop a unified framework for the measurement and valuation of health and productivity. Within this framework, we characterize evaluation functions allowing for compromises between the classical \textit{quality-adjusted life years} (QALYs) and its polar \textit{productivity-adjusted life years} (PALYs). Our framework and characterization results provide a new normative basis for the economic evaluation of health care interventions, as well as occupational health and safety policies, aimed to impact both health and productivity of individuals. 
\end{abstract}


\medskip{}

\medskip{}  \medskip{}

\newpage

\section{Introduction}
Few aspects concern human beings more than health. But resources are scarce and, as we face demographic changes, with increased demand from retirees and a constrained labour market due to shrinking working age share of the population, there is a pressing need to protect the health and productivity of the economically active population. Therefore, critical decisions on health care interventions, as well as occupational health and safety policies, have to be made constantly. The evidence from clinical trials and observational studies, in addition to assessments about the productivity consequences, are crucial to make those decisions. 
The purpose of this paper is to develop a unified framework for the measurement and valuation of outcomes of such programmes and policies. 

It is widely accepted that the health benefit a patient derives from a particular health care 
intervention can be defined according to two natural dimensions: quality of life and quantity of life. Pliskin et al. (1980) 
pioneered in multiattribute utility theory the axiomatic foundations of the so-called \textit{quality-adjusted life years} (in short, QALYs), which offer a straightforward procedure to combine the two natural dimensions. Together with the so-called \textit{disability-adjusted life years} (in short, DALYs), a primary focus of the landmark Global Burden of Disease studies, they are arguably the most frequently employed methodologies in the evaluation of health outcomes nowadays (e.g., Gold et al., 1996; Murray et al., 1997; Drummond et al., 2015).

An alternative to QALYs is the so-called \textit{productivity-adjusted life years} (in short, PALYs), which are calculated by multiplying a productivity index by years lived. The productivity index ranges from 0 (completely unproductive) to 1 (completely productive), and may take into consideration factors such as absence from work due to ill health (absenteeism), reduced productivity while at work (presenteeism) and premature exit from the workforce (e.g., Magliano et al., 2018; Ademi et al., 2021; Savira et al., 2021).

Economic evaluation of policies to improve occupational health and safety is one field of research where productivity outcome measures following the broader PALY idea are applied extensively (e.g., Tompa et al., 2009; Steel et al., 2018a, 2018b).
Similar to standard health economic evaluation, this involves measuring costs and effects of working environment interventions except that
many studies in this field concentrate on measuring changes in productivity as the most important
intervention effect while leaving health effects out (e.g., Noben et al., 2015; 
Finnes et al., 2022).
Productivity effects of interventions are often operationalised and measured for each individual employee as sickness absence and reduced productivity while the employees are at work
(presenteeism). Many studies are performed as a cost-benefit analysis from an employer's perspective where the purpose of the
analysis is to assess if the costs of an intervention can be covered entirely or to a degree through the
value of improved productivity (Grimani et al., 2018). However, from a societal perspective, it seems too narrow to focus
exclusively on productivity effects, as working environment interventions will often have wider
benefits including effects on employee health. 

Now, productivity effects are also often taken into account in the evaluation of health care itself (not just workplace safety). The traditional approach has been to net indirect earnings effects out of the cost of treatment (on the basis of the present value of the additional stream of lifetime income for an individual, possibly involving realistic situations such as involuntary unemployment) and keep the QALY outcome measure (e.g., Rice and Cooper, 1967; Koopmanschap et al., 1995; Sculpher, 2001). 
Another more recent approach, which is growing rapidly to the extent of becoming a new development in health economic evaluation, is precisely to switch from QALYs or DALYs to PALYs as the outcome measure. A sizable number of healthcare applications are lately endorsing this approach (see, for instance, Ademi et al. (2021) and the references cited therein). Nevertheless, PALYs as an outcome measure lack a proper theoretical justification. An aim of this paper is to fill that gap in this emerging literature by providing axiomatic foundations of this newly proposed outcome measure.

Furthermore, rather than replacing QALYs by PALYs, one could conceivably think about compromising between both, as they can each be seen as partial measures to evaluate outcomes of health care interventions or occupational health and safety policies. QALYs dismiss productivity, whereas PALYs dismiss quality-of-life (morbidity) concerns. More general evaluation functions encompassing both concerns would be more complete and thus more appropriate from a societal perspective. We also provide in this paper axiomatic foundations for evaluation functions allowing for compromises among QALYs and PALYs within a unified framework. In order to do that, we present a stylized model in which individuals are described by profiles of (representative) health states, productivity and lifetime spans. The aim is to derive measures to evaluate the distribution of those profiles in a population. And to do so by means of specific combinations of axioms that characterize different evaluation measures. 

Our approach builds upon the framework introduced in Hougaard et al. (2013). Therein, classical evaluation functions, such as the (time linear) QALY and HYE (acronym for \textit{healthy years equivalent}) measures are characterized axiomatically.\footnote{Moreno-Ternero et al. (2023) provide normative foundations for a general family of evaluations involving both QALYs and DALYs.} Our generalization of that framework allows us to characterize not only those evaluation functions but also others including a concern for productivity, such as the (time linear) PALY evaluation function, and also appealing hybrids between QALYs and PALYs.\footnote{Our approach will thus be somewhat reminiscent of Keeney and Smith (2005), who developed a combined health-and-consumption adjusted life year concept, albeit for individual (rather than social) decision making. See also Bleichrodt and Quiggin (1999).} We shall refer to the simplest one as \textit{productivity-and-quality-adjusted life years} (in short, PQALYs). We shall characterize other hybrids as well, combining QALYs with PALYs or PQALYs, in an additive way.
We shall also characterize even more general evaluations functions, around the notion of \textit{healthy productive years equivalent} we introduce in this setting. 

We conclude this introduction stressing that our model treats health and productivity as different individual attributes. 
In doing so, we obviously depart from the literature that considers only one of them, but also from the simplistic assumption that both concepts are perfectly correlated (which would allow to use a reduced model). The precise relationship between health and productivity is complex and the anticipated correlation might actually be positive or negative, depending on the viewpoint. For instance, Tompa (2002) reviews a literature which suggests that ``chronic and acute physical and mental conditions, as well as health-related behaviours, explain a significant portion of sickness absence (a proxy for productivity)". 
On the other hand, Hummels et al., (2023) recently studied the effect of rising workload on health and found that it wears down an individual worker's health capital, leading to an increased likelihood of sickness. 
This is also in line with the classical human capital model of the demand for health (e.g., Grossman, 1972, 2000), which concludes that health does not affect productivity (it affects an individual's annual salary but not the individual's hourly wage).

The rest of the paper is organized as follows. 
In Section \ref{sec:prelim}, we introduce the framework and the basic common axioms that all our evaluation functions will satisfy. In Section \ref{sec:time linear}, we characterize the focal (and somewhat polar) evaluation functions QALYs and PALYs. In Section \ref{compromises}, we characterize classes of evaluation functions which compromise among the previous two. 
In Section \ref{general}, we characterize more general functional forms, that evolve around the notion of \textit{healthy productive years equivalent}. In Section \ref{sec:discussion}, we discuss our contribution with a special emphasis on the choice among the different evaluation functions we characterize. Finally, in Section \ref{sec:final remarks}, we provide some concluding remarks providing further connections to related literature and pointing out possible extensions of our work. For a smooth passage, we defer all proofs to the Appendix.

\section{Preliminaries}\label{sec:prelim}
Let a population consisting of $n$ individuals be identified with the set $N=\{1,...,n\}$. Each individual $i \in N$ is described by a profile, formalized by a triple $d_i=(a_i, p_i, t_i)$, where $a_i\in A$ is a health state, $p_i \in  [0, 1]$ is the productivity level, and $t_i \in \Bbb{R}_+ = [0, \infty)$ is time. 

We can think of the health state $a_i$ as a chronic or representative health state over time.\footnote{Our analysis also allows to consider time-varying health, as we discuss in Section 7.} 
We assume that there exists a health state $a_\ast$, referred to as `full health', which is considered at least as good as any other health state. To keep our analysis as general as possible, we make no further assumptions regarding the domain of health states $A$.  

The productivity $p_i$ is measured by any chosen indicator. For instance, it can be an indicator of absence from work (e.g., number of sick days per year for a person). Note that such an indicator may reflect productivity and contributions to society in a broad sense. For example, work may include both labour market activities and domestic work. Alternatively, a measure could be chosen that reflects the value of the work contributed by the individuals, as would (very roughly) be the case if measured by e.g., (relative) earnings (e.g., Steel et al., 2018b).\footnote{Thus, if the policy maker wants to take into account an estimate of individual productivity in monetary means for society, this could be done by letting $p_i$ be the annual earnings measured as a fraction of the highest earnings in society.} However, our primary interpretation of productivity will be the broad one, as taking into account the monetary value of individuals (to society) is ethically questionable, as well as potentially more difficult to measure. 

Finally, there are two plausible interpretations of time in our model. On the one hand, it could be identifying the individual total lifetime. On the other hand, it could be identifying incremental individual lifetime from a given status quo up to the end of life or retirement. 

Let $d=(d_1,...,d_n)$ denote a distribution of individual profiles, as described above, and let $D$ denote the set of possible distributions. We now give an example in which two hypothetical distributions are presented. We shall return to this example later in the text several times to illustrate how these two hypothetical distributions can be (relatively) evaluated,  by means of various evaluation functions we consider. 

\begin{example} Consider the following two distributions, 
involving five individuals each (that could be interpreted as representative agents of five different groups). 

In the first distribution ($d^{\Delta}$), all individuals are experiencing full health. The first one is also experiencing maximum productivity as well as forty years of lifetime (until retirement, or the end of life), i.e., $d^{\Delta}_1=(a_\ast,1,40)$. The second individual is experiencing $50\%$ of maximum productivity and forty years of lifetime, i.e., $d^{\Delta}_2=(a_\ast,0.5,40)$. The third individual is experiencing zero productivity and forty years of lifetime, i.e., $d^{\Delta}_3=(a_\ast,0,40)$. The fourth individual is experiencing $50\%$ of maximum productivity and ten years of lifetime, i.e., $d^{\Delta}_4=(a_\ast,0.5,10)$. The last individual is experiencing zero productivity and lifetime, i.e., $d^{\Delta}_5=(a_\ast,0,0)$.

In the second distribution ($d^{\Lambda}$), the first individual is also experiencing full health and maximum productivity as well as forty years of lifetime, i.e., $d^{\Lambda}_1=d^{\Delta}_1=(a_\ast,1,40)$. The remaining individuals are experiencing a worse health state, which we denote by $a$.\footnote{Reflecting, for instance, some problems in mobility, self-care, usual activities, pain/discomfort and anxiety/depression to consider the standard dimensions in the EuroQol system (e.g., Rabin and Charro, 2001).} The second individual is also experiencing maximum productivity and forty years of lifetime, i.e., $d^{\Lambda}_2=(a,1,40)$. The third individual is experiencing $50\%$ of maximum productivity and forty years of lifetime, i.e., $d^{\Lambda}_3=(a,0.5,40)$. The fourth individual is experiencing $50\%$ of maximum productivity and ten years of lifetime, i.e., $d^{\Lambda}_4=(a,0.5,10)$. The last individual is experiencing zero productivity and lifetime (until retirement), i.e., $d^{\Lambda}_5=(a,0,0)$.

In summary, 
$$d^{\Delta}=\left(\begin{array}{ccc}d^{\Delta}_1\\ 
d^{\Delta}_2\\
d^{\Delta}_3\\ 
d^{\Delta}_4\\
d^{\Delta}_5 \end{array}\right) 
=
\left(\begin{array}{ccc}a_\ast & 1 & 40\\ 
a_\ast & 0.5 & 40\\
a_\ast & 0 & 40\\ 
a_\ast & 0.5 & 10\\
a_\ast & 0 & 0 \end{array}\right); \quad 
d^{\Lambda}=\left(\begin{array}{ccc}d^{\Lambda}_1\\ 
d^{\Lambda}_2\\
d^{\Lambda}_3\\ 
d^{\Lambda}_4\\
d^{\Lambda}_5 \end{array}\right) 
=
\left(\begin{array}{ccc}a_\ast & 1 & 40\\ 
a & 1 & 40\\
a & 0.5 & 40\\ 
a & 0.5 & 10\\
a & 0 & 0\end{array}\right).$$
\end{example}
\medskip

The preferences of a social planner (or social preferences) over distributions is given by $\succsim$.  
A distribution evaluation function (\textit{evaluation function}, in short) is a real-valued function $E: D \rightarrow \mathbb{R}$. We say that $E$ represents $\succsim$ if \[E[d]
\ge E[d'] \Leftrightarrow d \succsim d'.\] Note that if $E$ represents $\succsim
$ then any strictly increasing transformation of $E$ would also do so.

An evaluation function $E$ may be interpreted as an effect measure if it is used for the economic evaluation of health care or working environment interventions.

In what follows, we present some basic axioms for social preferences in the current context, that will be common to all the evaluation functions we consider in this paper.\footnote{The axioms are an adaptation of those in Hougaard et al. (2013) and Moreno-Ternero et al. (2023) for social preferences over distributions to the present enriched model to account for productivity.}

\subsection{COMMON axioms}

In this section, we present a set of seven axioms that forms the necessary conditions for the theorems presented in the remaining sections of this paper. These are termed the COMMON axioms. The axioms reflect basic principles adapted to our framework that are widely accepted in economics. In the following sections, additional axioms are presented, which together with the COMMON axioms close the characterizations of the evaluation functions we highlight. 

The first three COMMON axioms apply to all three attributes in the same way; whereas the latter four COMMON axioms introduce some conditions on time which distinguish it from the other two attributes. 

The first axiom, \textit{anonymity}, reflects the principle of impartiality, with a long tradition in the theory of justice (e.g., Moreno-Ternero and Roemer, 2006). It
says that the identities of individuals do not matter in the evaluation of distributions. Formally, permuting triples does not alter preferences. 

\vskip3mm\noindent \textbf{ANON}: $(d_1,...,d_n)\sim (d_{\pi(1)},...,d_{\pi(n)})$ for each $d\in D$, and each bijection $\pi: N\to N$.
\medskip

The second axiom, \textit{separability}, also has a long tradition of use in models of welfare economics (e.g., Moulin, 1988). It says that if the profiles change only for a subgroup of individuals, then the relative evaluation of the two distributions should only depend on this subgroup.\footnote{In Example 1, the axiom implies that replacing individual 1 in both distributions by an individual with another profile would not alter the preferences between both distributions.} 

\vskip3mm\noindent \textbf{SEP}: For each pair $d,d'\in D$, and each $S\subset N$, $(d_S,d_{N\setminus S}) \succsim (d'_S,d_{N\setminus S}) \Leftrightarrow (d_S,d'_{N\setminus S}) \succsim (d'_S,d'_{N\setminus S})$. 

\medskip

The third axiom, \textit{continuity}, is the adaptation of a standard operational condition to our context. It says that small changes in productivity or life years should only produce small changes in the evaluation of the distribution.\footnote{Note that this axiom is not exactly treating the three attributes in the same way, as the first one (health) is not endowed with a mathematical structure. Its nature is nevertheless somewhat different to the nature of the axioms in the second group we present next.} 

\vskip3mm\noindent \textbf{CONT}: Let $d,d'\in D$, and $d^{(k)}$ be a sequence such that for each $i \in N$, $d_i^{(k)}=(a_i,p_i^{(k)}, t_i^{(k)}) \rightarrow (a_i,p_i, t_i)=d_i$. If $d_i^{(k)} \succsim d'$ for each $k$ then $d \succsim d'$, and if $d' \succsim d_i^{(k)}$ for each $k$, then $d' \succsim d$. 

\medskip

We then move to the second group of COMMON axioms. First, the \textit{social zero condition}, which is reminiscent of a well-known condition for individual utility functions on health (e.g., Miyamoto et al., 1998). It says that if an individual has zero lifetime, then the health state and productivity of that individual does not influence the evaluation of the distribution.\footnote{In Example 1, this implies that changing the health or productivity of individual 5 in each distribution would render the resulting distributions equally valued.}


\vskip3mm\noindent \textbf{ZERO}: For each $d \in D$ and each $i \in N$ such that $t_i = 0$, and each $a^{\prime}_{i}\in A$, and each $p_i' \in [0,1]$, $d\sim [d_{N\setminus \{i\}}, (a_i', p_i', 0)]$. 

\medskip



The next axiom, \textit{full health and productivity superiority}, formalizes in our context a certain form of the Pareto principle of optimality. It says that replacing an individual's health state by that of full health, or increasing productivity to its maximum, cannot worsen the evaluation of the distribution. 


\vskip3mm\noindent \textbf{FHPS}: For each $d\in D$ and each $i\in N$, $[(a_\ast,p_i, t_i),d_{N\setminus \{i\}}] \succsim d$ and $[(a_i,1, t_i),d_{N\setminus \{i\}}] \succsim d$.

\medskip
The following axiom, \textit{lifetime monotonicity at full health and productivity}, says that if an individual enjoys full health and maximum productivity then increasing life years is strictly better.


\vskip3mm\noindent \textbf{LMFHP}: For each $d \in D$ and each $i \in N$, such that $(a_i,p_i)=(a_\ast,1)$ and each $t_i > t_i'$, $[(a_\ast,1, t_i), d_{ N\setminus \{i\}}] \succ [(a_\ast,1,t_i'), d_{ N\setminus \{i\}}]$. 

\medskip

The last one, \textit{positive lifetime desirability}, requires special treatment as it says that the distribution does not worsen if any individual moves from zero lifetime to positive lifetime, when the health state and productivity is kept fixed. This implies, in particular, that health states worse than death are excluded. 


\vskip3mm\noindent \textbf{PLD}: For each $d \in D$ and each $i \in N$, $ d \succsim [(a_i,p_i,0), d_{N\setminus \{i\}}]$.



\section{QALYs and PALYs}\label{sec:time linear}

Economic evaluations of health care interventions often focus on health improvements of individuals as the main outcome of interest (e.g., Drummond et al., 2015). In contrast, economic evaluations of interventions to improve the working environment have a stronger focus on improving productivity, often measured as reductions in sickness absence and enhancing productivity while at work (e.g., Tompa et al., 2008). In this section, we investigate the assumptions underlying such evaluation functions with a narrow focus. We shall move in the next sections to investigate more general evaluation functions with a broader scope.  


We start with the classical \textit{(aggregated time-linear) QALY} evaluation function. It evaluates distributions by
means of the unweighted aggregation of individual QALYs in society,
or, in other words, by the weighted (through health levels)
aggregate time span the distribution yields.  
Formally,
\begin{equation}  \label{eqQALY}
E^q[d_1,\dots,d_n]= E[(a_1,p_1,t_1),\dots, (a_n,p_n,t_n)]=\sum_{i=1}^n q(a_i)t_i,
\end{equation}
where $q:A\to[0,1]$ is a function satisfying $0 \le q(a_{i})\le
q(a_{\ast})=1$, for each $a_{i}\in A$.

The unweighted aggregation of individual QALYs, as specified in (\ref{eqQALY}), is the preferred evaluation function in the majority of cost-utility analyses performed for health care interventions (e.g., Gold et al., 1996; Drummond et al., 2015).

This evaluation function ignores productivity. More precisely, it satisfies the following axiom, \textit{productivity independence}, which states that for a fixed health state and lifetime, the productivity is irrelevant for the evaluation of the distribution.

\vskip3mm\noindent \textbf{PI}: For each $d \in D$, each $i \in N$ and $p_i'\not =p_i$, $ [(a_i,p_i,t_i), d_{N\setminus\{i\}}] \sim [(a_i,p_i',t_i), d_{N\setminus\{i\}}]$. 
\bigskip

It also satisfies the \textit{time invariance at common health and full productivity} axiom, which states that for two individuals at common health and maximum productivity, extra life years are interchangeable. That is, it does not matter to the social planner which individual (among those with common health and maximum productivity) gets extra life years.

\vskip3mm\noindent \textbf{TICHFP}: For each $d\in D$, each pair $i,j\in N$ with $a_i=a_j=a$ and $p_i=p_j=1$, and each $c>0$, $$[(a,1,t_i+c),(a,1,t_j), d_{ N\setminus \{i,j\}}] \sim [(a,1,t_i),(a,1,t_j+c), d_{ N\setminus \{i,j\}}].$$

Our first result states the QALY evaluation function is characterized by the combination of the previous two axioms and the COMMON axioms.\footnote{Theorem \ref{QALY} is the extension to our setting of the counterpart results in Hougaard et al. (2013) and Moreno-Ternero et al. (2023).}


\begin{theorem} \label{QALY}
The following statements are equivalent:
\begin{enumerate}
\item $\succsim $ is represented by a QALY evaluation function (\ref{eqQALY}). 
\item $\succsim $ satisfies COMMON, PI, and TICHFP. 
\end{enumerate}
\end{theorem}

A possible interpretation of Theorem \ref{QALY} could be a situation where a social planner decides to evaluate and prioritise among a set of interventions using the QALY evaluation function $(\ref{eqQALY})$. This social planner will thus agree to the value choices that productivity changes among individuals do not matter for the evaluation of interventions (axiom PI) and that an extension of life of a specific length of any two individuals is of equal value as long as they enjoy a common health state and maximum productivity (axiom TICHFP). In addition, the social planner subscribes to a set of assumptions that are largely technical and uncontroversial (the COMMON axioms).\footnote{As all theorems in this paper require the COMMON axioms, these will not be mentioned again in the interpretations of the remaining theorems.}

A counterpart axiom of \textit{productivity independence} is \textit{health independence}, which states that, for fixed productivity and lifetime, the health state of an individual is irrelevant for the evaluation.  


\vskip3mm\noindent \textbf{HI}: For each $d \in D$, each $i \in N$ and $a_i'\not =a_i$, 
$$[(a_i,p_i,t_i), d_{ N\setminus \{i\}}] \sim [(a'_i,p_i,t_i), d_{ N\setminus \{i\}}].$$

And, likewise, a counterpart of \textit{time invariance at common health and full productivity} is \textit{time invariance at full health and common productivity}, which states that for two individuals at full health and common productivity, it does not matter to the social planner who receives the extra life years.


\vskip3mm\noindent \textbf{TIFHCP}: For each $d\in D$, each pair $i,j\in N$ with $a_i=a_j=a_{\ast}$ and $p_i=p_j=p$, and each $c>0$,
$$[(a_{\ast},p,t_i+c),(a_{\ast},p,t_j), d_{ N\setminus \{i,j\}}] \sim [(a_{\ast},p,t_i),(a_{\ast},p,t_j+c), d_{ N\setminus \{i,j\}}].$$

As the next result states, if the previous two axioms replace their counterparts at Theorem \ref{QALY}, we characterize the following \textit{generalized PALY} evaluation function, which evaluates distributions by means of the aggregation of individual PALYs in society, when submitted first to a continuous function ($v$). 
Formally,
\begin{equation}\label{eqgPALY}
 E^{vp}[d_1,\dots,d_n]= E[(a_1,p_1,t_1),\dots, (a_n,p_n,t_n)]=\sum_{i=1}^n v(p_i)t_i,
\end{equation}
where $v:[0,1]\to[0,1]$ is a continuous function satisfying $0 \le v(p_{i})\le
v(1)=1$, for each $p_{i}\in [0,1]$.

\begin{theorem} \label{gPALY}
The following statements are equivalent:
\begin{enumerate}
\item $\succsim $ is represented by a generalized PALY evaluation function (\ref{eqgPALY}). 
\item $\succsim $ satisfies COMMON, HI, and TIFHCP. 
\end{enumerate}
\end{theorem}

The generalized PALY evaluation function is a counterpart of the QALY evaluation function. As such, neither the $v$ function nor the $q$ function (in their respective functional forms) has a monotonic structure. This makes sense in the latter case because the domain of health states $A$ does not have a mathematical structure. But the domain of productivity levels is naturally ordered and, therefore, it would make sense to impose $v$ a non-decreasing structure. The following axiom will guarantee such a feature as a byproduct. 

The axiom \textit{productivity invariance at full health and common time} states that, for any two individuals with common lifetime and full health, it makes no difference who gains in productivity.

\vskip3mm\noindent \textbf{PIFHCT}: For each $d\in D$, each pair $i,j\in N$ with $a_i=a_j=a_\ast$ and $t_i=t_j=t$, and each $c>0$ such that $p_i+c,p_j+c \leq 1$, $$[(a_\ast,p_i+c,t),(a_\ast,p_j,t), d_{ N\setminus \{i,j\}}] \sim [(a_\ast,p_i,t),(a_\ast,p_j+c,t), d_{ N\setminus \{i,j\}}].$$

As the next result states, adding this axiom to those in Theorem \ref{gPALY} we characterize the \textit{affine PALY} evaluation function, which evaluates distributions by means of the aggregation of individual PALYs in society, when submitted first to an affine and non-decreasing function. 
Formally,
\begin{equation}\label{affPALY}
 E^{ap}[d_1,\dots,d_n]= E[(a_1,p_1,t_1),\dots, (a_n,p_n,t_n)]=\sum_{i=1}^n (\alpha p_i+(1-\alpha)) t_i,
\end{equation}
where $\alpha\in [0,1]$.


\begin{theorem}\label{a-PALY} 
    The following statements are equivalent:
\begin{enumerate}
\item $\succsim $ is represented by an affine PALY evaluation function (\ref{affPALY}).
\item $\succsim $ satisfies COMMON, HI, TIFHCP, and PIFHCT. 
\end{enumerate}
\end{theorem}

The previous families are obvious generalizations of the focal \textit{linear PALY} evaluation function, which evaluates distributions by
means of the unweighted aggregation of individual PALYs in society,
or, in other words, by the weighted (through productivity levels)
aggregate time span the distribution yields.  
Formally,
\begin{equation}\label{eqPALY}
 E^p[d_1,\dots,d_n]= E[(a_1,p_1,t_1),\dots, (a_n,p_n,t_n)]=\sum_{i=1}^n p_it_i.
\end{equation}

The unweighted aggregation of individual PALYs as specified in (\ref{eqPALY}) is a common evaluation function used for economic evaluation of working environment interventions including, for example, Finnes et al. (2022), who measured sickness absence days as effect measure.

This evaluation function is characterized when we add the \textit{time independence for unproductive individuals} axiom, which states that if an individual has zero productivity, then the lifetime of that individual does not influence the evaluation of the distribution.

\vskip3mm\noindent \textbf{TIUP}: For each $d \in D$ and each $i \in N$ such that $p_i = 0$, and each $t^{\prime}_{i}\in \Bbb{R}_+$, $d\sim [d_{N\setminus \{i\}}, (a_i, 0, t^{\prime}_{i})]$. 

 
\begin{theorem}\label{PALY} 
    The following statements are equivalent:
\begin{enumerate}
\item $\succsim $ is represented by a PALY evaluation function (\ref{eqPALY}).
\item $\succsim $ satisfies COMMON, HI, TIFHCP, PIFHCT, and TIUP. 
\end{enumerate}
\end{theorem}

According to Theorem \ref{PALY}, a social planner wishing to conduct an economic evaluation of working environment interventions using the PALY evaluation function (\ref{eqPALY}) will also hold a number of specific values in the process of priority setting. These value choices include, for example, that changes in health among individuals following an intervention (axiom HI) and that increases in lifetime among unproductive individuals (axiom TIUP) both have no influence on the choice of intervention.

We conclude this section applying the evaluation functions characterized in this section to the distributions from Example 1. More precisely, we summarize the computations in the next table. We infer from there that the first distribution is preferred from the viewpoint of QALYs, whereas the second distribution is preferred from the viewpoint of the PALYs-based evaluation functions.\footnote{Recall that $\alpha,v(p),q(a)\in[0,1]$.} 
\begin{center}
\begin{tabular}{|l|l|l|l|}
\hline
 & \qquad\quad $E[d^{\Delta}]$ & \quad\,\, $E[d^{\Lambda}]$ & \quad\,\,$\succsim$\\ \hline
$E^q$ & $130$ & $ 40+90q(a)$& $d^{\Delta}\succsim d^{\Lambda}$ \\ \hline 
$E^p$ & $ 65$ & $ 105$& $d^{\Lambda}\succsim d^{\Delta}$\\ \hline 
$E^{ap}$ & $ 130-65\alpha$ & $ 130-25\alpha$& $d^{\Lambda}\succsim d^{\Delta}$ \\ \hline 
$E^{vp}$ & $ 40+50v(0.5)+40v(0)$ & $ 80+50v(0.5)$& $d^{\Lambda}\succsim d^{\Delta}$ \\ \hline 
\end{tabular}
\end{center}

\section{Compromising between QALYs and PALYs}\label{compromises}
As mentioned above, the evaluation functions characterized in the previous section ignore one dimension of our model. In other words, they all rely on a very demanding axiom of (productivity or health) independence. The purpose of this section is to dismiss those axioms, while obtaining characterizations of natural compromises between those focal (albeit polar) evaluation functions. 

For instance, the \textit{productivity-and-quality-adjusted life years} (PQALY) evaluation function evaluates distributions by means of the weighted (through productivity and health levels) aggregate time span the distribution yields, so that health, productivity and lifespan of individuals enter the evaluation function multiplicatively.  
Formally,
\begin{equation}\label{eqM-PQALY}
E^{pq}[d_1,\dots,d_n]= E[(a_1,p_1,t_1),\dots, (a_n,p_n,t_n)]=\sum_{i=1}^n q(a_i)p_it_i,
\end{equation}
where $q:A\rightarrow [0,1]$ is a health state quality weight satisfying $0\le q(a_i) \le q(a_\ast)= 1$ 
for each $a_i\in A$.

As the next result states, this evaluation function is characterized when we dismiss health independence in the previous result (characterizing PALYs), and strengthen the other two independence axioms to consider the following ones. 

First, \textit{time invariance at common health and productivity}, which states that for two individuals at common health and productivity, it does not matter to the social planner who receives the extra life years.


\vskip3mm\noindent \textbf{TICHP}: For each $d\in D$, each pair $i,j\in N$ with $a_i=a_j=a$ and $p_i=p_j=p$, and each $c>0$,
$$[(a,p,t_i+c),(a,p,t_j), d_{ N\setminus \{i,j\}}] \sim [(a,p,t_i),(a,p,t_j+c), d_{ N\setminus \{i,j\}}].$$

Second, \textit{productivity invariance at common health and time}, which says that for two individuals at common health and time, it does not matter who gains in productivity.

\vskip3mm\noindent \textbf{PICHT}: For each $d\in D$, each pair $i,j\in N$ with $a_i=a_j=a$ and $t_i=t_j=t$, and each $c>0$ such that $p_i+c,p_j+c \leq 1$,
$$[(a,p_i+c,t),(a,p_j,t), d_{ N\setminus \{i,j\}}] \sim [(a,p_i,t),(a,p_j+c,t), d_{ N\setminus \{i,j\}}].$$

\begin{theorem}\label{thm:M-PQALY} 
The following statements are equivalent:
\begin{enumerate}
 \item $\succsim $ is represented by a PQALY evaluation function (\ref{eqM-PQALY}).
\item $\succsim $ satisfies COMMON, TICHP, PICHT, and TIUP. 
\end{enumerate}
\end{theorem}

A social planner may view both health effects and productivity effects as important outcomes of interventions and may therefore choose an evaluation function like the PQALY (\ref{eqM-PQALY}), where health status and productivity of individuals enter the evaluation function multiplicatively. This implies according to Theorem \ref{thm:M-PQALY} that one of the values applied by the social planner in this situation is that increases in lifetime among unproductive individuals following an intervention (axiom TIUP) will not increase the desirability of that intervention. The social planner subscribes to two further values regarding invariance between different effect components where the first states that if an intervention leads to extra life years, it does not matter to the social planner which particular individual (among individuals with the same level of health and productivity) receives these extra life years (axiom TICHP). The second value specifies that if an intervention leads to improved productivity, it does not matter which particular individual (among individuals with the same level of health and lifespan) is able to perform better in the workplace (axiom PICHT).

Our next result states that dismissing the TIUP axiom in the previous statement we obtain the following alternative intriguing compromise (dubbed 
\textit{QALY-PQALY}) which evaluates distributions by means of a convex combination of the QALYs and PQALYs the distribution yields.  
Formally,
\begin{equation}\label{eqA-PQALY}
E^{\delta}[d_1,\dots,d_n]= E^{\delta}[(a_1,p_1,t_1),\dots, (a_n,p_n,t_n)]=\delta\sum_{i=1}^n q(a_i)t_i+(1-\delta)\sum_{i=1}^n r(a_i)p_it_i,
\end{equation}
where $q,r:A\rightarrow [0,1]$ are health state quality weight functions satisfying $0\le q(a_i) \le q(a_\ast)= 1$ and $0\le r(a_i) \le r(a_\ast)= 1$ 
for each $a_i\in A$, and $\delta\in[0,1]$.

The QALY-PQALY evaluation function (\ref{eqA-PQALY}) divides the evaluation exercise into two separate tasks; namely the valuation of the pure health effects among individuals following an intervention, plus the valuation of productivity-and-quality adjusted lifespans of individuals. The parameter $\delta$ measures the relative importance that the social planner puts on pure health effects and productivity-and-quality adjusted life years resulting from an intervention.

\begin{theorem}\label{thm:A-PQALY} 
The following statements are equivalent:
    \begin{enumerate}
     \item $\succsim $ is represented by a QALY-PQALY evaluation function (\ref{eqA-PQALY}).
    \item $\succsim $ satisfies COMMON, TICHP, and PICHT. 
    \end{enumerate}
\end{theorem}

Similar to the previous evaluation function, a social planner choosing the QALY-PQALY evaluation function (\ref{eqA-PQALY}) considers both health effects and productivity effects as important outcomes of interventions. In contrast to the previous theorem, Theorem \ref{thm:A-PQALY} implies a rejection of axiom TIUP according to which improvements in lifetime among unproductive individuals following an intervention do not increase the desirability of that intervention. Apart from that, the social planner subscribes to the same two axioms regarding invariance between different effect components as above (TICHP and PICHT).

The previous family of evaluation functions (\ref{eqA-PQALY}) includes a natural sub-family of evaluation functions (\textit{QALY-PALY}) that evaluate distributions by means of the convex combinations of the QALYs and PALYs that the distribution yields. Formally, 
\begin{equation}\label{QALY-PALY}
E^{\sigma}[d_1,\dots,d_n]= E^{\sigma}[(a_1,p_1,t_1),\dots, (a_n,p_n,t_n)]=\sigma\sum_{i=1}^n q(a_i)t_i+(1-\sigma)\sum_{i=1}^n p_it_i,
\end{equation}
where $q:A\rightarrow [0,1]$ is a health state quality weight function satisfying $0\le q(a_i) \le q(a_\ast)= 1$, 
for each $a_i\in A$, and $\sigma\in[0,1]$.

Health effects and productivity effects enter the QALY-PALY evaluation function (\ref{QALY-PALY}) additively so that these two effects are measured and valued independently of each other. The parameter $\sigma$ indicates the relative importance that the social planner places on health and productivity effects of individuals respectively.

This family of evaluation functions (\ref{QALY-PALY}) is characterized when strengthening the \textit{productivity invariance at common health and time} axiom to \textit{productivity invariance at common time}, which says that for two individuals with common lifespan (and irrespective of their health status), it does not matter which of the two individuals improve their productivity.

\vskip3mm\noindent \textbf{PICT}: For each $d\in D$, each pair $i,j\in N$ with $t_i=t_j=t$, and each $c>0$ such that $p_i+c,p_j+c \leq 1$,
$$[(a_i,p_i+c,t),(a_j,p_j,t), d_{ N\setminus \{i,j\}}] \sim [(a_i,p_i,t),(a_j,p_j+c,t), d_{ N\setminus \{i,j\}}].$$

\begin{theorem}\label{thm:QALY-PALY} 
The following statements are equivalent:
\begin{enumerate}
 \item $\succsim $ is represented by a QALY-PALY evaluation function (\ref{QALY-PALY}).
\item $\succsim $ satisfies COMMON, TICHP and PICT. 
\end{enumerate}
\end{theorem}

Theorem \ref{thm:QALY-PALY} also implies a rejection of axiom TIUP, where improvements in lifetime among unproductive individuals do not matter for the choice among interventions. Further, the social planner agrees to the value that if an intervention leads to extra life years, it does not matter to the social planner which particular individual (among individuals with the same level of health and productivity) receives these extra life years (axiom TICHP). Finally, if an intervention leads to improved productivity, it does not matter to the social planner which particular individual (among individuals with the same level of lifespan but not necessarily the same health status) has improved productivity (axiom PICT).

We also conclude this section applying the evaluation functions characterized in this section to the distributions from Example 1. More precisely, we summarize the computations in the next table. We infer from there that a distribution is preferred to the other one for a certain range of the parameters defining the families of evaluation functions we are considering in this section. For instance, if we assume $q(a)=r(a)=0.5$, then $E^\delta(d^{\Delta})<E^\delta(d^{\Lambda})$ for all members of the family (\ref{eqA-PQALY}). On the other hand, if we assume $q(a)=r(a)=0$, then $E^\delta(d^{\Delta})>E^\delta(d^{\Lambda})$ for all members of the family (\ref{eqA-PQALY}). 
Finally, if we assume $q(a)=r(a)=0.4$, then $E^\delta(d^{\Delta})<E^\delta(d^{\Lambda})$ if and only if $\delta<\frac{1}{3}$. As for the family (\ref{QALY-PALY}), $E^\sigma(d^{\Delta})<E^\sigma(d^{\Lambda})$ if and only if $\sigma<\frac{4}{13-9q(a)}$.
\begin{center}
\begin{tabular}{|l|l|l|l|}
\hline
 &\,\,\, $E[d^{\Delta}]$ & \qquad\qquad\quad $E[d^{\Lambda}]$ & \qquad\qquad\qquad \qquad$\succsim$ \\ \hline 
$E^{pq}$ & $65$ & $ 40+ 65q(a) $&\, $d^{\Delta}\succsim d^{\Lambda} $ if and only if $q(a)\le\frac{5}{13} $ \,\\ \hline
$E^\delta$ & $65 (1+\delta)$ & $ 40+90\delta q(a)+65 (1-\delta)r(a)$&\, Depends on $q,r,\delta$ \,\\ \hline
$E^\sigma$ & $65(1+\sigma)$ & $ 105-65 \sigma+90 q(a)\sigma $&\, $d^{\Delta}\succsim d^{\Lambda} $ if and only if $\sigma\ge\frac{4}{13-9q(a)}$ \, \\ \hline
\end{tabular}
\end{center}

\section{More general evaluation functions}\label{general}
We have presented some central compromises to evaluate both health and productivity aspects of distributions in the previous section. We can even be more flexible to accommodate other (more general) compromises. As a matter of fact, those compromises considered above can be seen as members of a general family of evaluation functions that evaluate distributions by the weighted aggregation of lifetimes the distribution yields, where the weight is a function of health and productivity.  
Formally,
\begin{equation}\label{tw-HPYE}
E^w[d_1,\dots,d_n]= E[(a_1,p_1,t_1),\dots, (a_n,p_n,t_n)]=\sum_{i=1}^n w(a_i,p_i)t_i,
\end{equation}  
where  $ w :A\times [0,1] \rightarrow [0,1]$ is a continuous function in its second variable and such that $0\le w(a,p)\le w(a_{\ast},p), w(a,1)\le w(a_{\ast},1)=1$ for each $(a,p)\in A\times[0,1]$. 
\bigskip

Health effects and productivity effects enter the evaluation function (\ref{tw-HPYE}) via the general function $w$, which sets the weights for individual lifetimes in the ensuing aggregation. This permits to accommodate a wide variety of options to account for the interaction between health effects and productivity effects. In particular, the weight function $w$ can fully ignore one or the other, thus giving rise to PALYs (generalized or not) and QALYs respectively. But it can also be a multiplicative function, thus giving rise to the PQALY evaluation function; or a linear function, thus giving rise to the QALY-PALY evaluation function. Finally, it can also be a semilinear function, thus giving rise to the QALY-PQALY evaluation function.

But the family of evaluation functions (\ref{tw-HPYE}) can also accommodate other evaluation functions that have not been introduced above. For instance, suppose that $w$ is a semimultiplicative function in which productivity enters via a power function, whereas health enters via QALYs, i.e., $ w(a_i,p_i)=q(a_i)p_i^{\gamma}$, for each $(a_i,p_i)\in A\times[0,1]$, where $\gamma\in (0,1)$. 

\begin{equation*}\label{gamma}
    E^{\gamma}[d_1,\dots,d_n]= E[(a_1,p_1,t_1),\dots, (a_n,p_n,t_n)]=\sum_{i=1}^n q(a_i)p_i^{\gamma}t_i,
    \end{equation*}  
    where $q:A\rightarrow [0,1]$ is a health state quality weight function satisfying $0\le q(a_i) \le q(a_\ast)= 1$, 
    for each $a_i\in A$, and $\gamma\in (0,1)$. 

Note that the previous evaluation function is formalizing a concern for the dispersion of productivity levels (as it is a concave function of those levels).

The above is somewhat reminiscent of the focal welfare function within the literature on life-cycle preferences over consumption and health status. Therein, multiplicative separability from consumption and health is typically assumed.\footnote{Bleichrodt and Quiggin (1999) show that the condition is necessary for lifetime utility maximisation to be consistent with cost-effectiveness analysis and so equivalent to cost-benefit analysis.} 
In our case, we have productivity instead of consumption, but multiplicative separability can also be naturally obtained from the general family of evaluation functions (\ref{tw-HPYE}).

It turns out that the general family of evaluation functions (\ref{tw-HPYE}) is characterized when combining the COMMON set of axioms and \textit{time invariance at common health and productivity}.

\begin{theorem}\label{thm:tw-HPYE}
    The following statements are equivalent:
    \begin{enumerate}
     \item $\succsim $ is represented by an evaluation function (\ref{tw-HPYE}).
    \item $\succsim $ satisfies COMMON and TICHP. 
    \end{enumerate}
\end{theorem}

Theorem \ref{thm:tw-HPYE} implies that if the social planner endorses the view that if an intervention leads to extra life years, it does not matter which particular individual (among individuals with the same level of health and productivity) receives these extra life years (axiom TICHP), then the evaluation will be via a weighted aggregation of the lifetimes the intervention yields. And the weight for each individual lifetime will be obtained via a general function of the health and productivity levels they face.


A weaker axiom than \textit{time invariance at common health and productivity},  is \textit{time invariance at full health and productivity}, which states that extra years can be interchangeable among individuals with full health and maximum productivity. 

\vskip3mm\noindent \textbf{TIFHP}: For each $d\in D$, each pair $i,j\in N$ with $a_i=a_j=a_\ast$ and $p_i=p_j=1$, and each $c>0$, $$[(a_\ast,1, t_i+c), d_{ N\setminus \{i\}}] \sim [(a_\ast,1,t_j+c), d_{ N\setminus \{j\}}].$$


If we replace \textit{time invariance at common health and productivity} by \textit{time invariance at full health and productivity} we characterize a more general family of evaluation functions, which extend to this context the notion of \textit{healthy years equivalent} (e.g., Mehrez and Gafni, 1989). 


More precisely, the \textit{healthy productive years equivalent} (\textit{HPYE}) evaluation function evaluates distributions by the unweighted aggregation of HPYEs the distribution yields.  
Formally,
\begin{equation}\label{eqlinHPYE}
E^f[d_1,\dots,d_n]= E[(a_1,p_1,t_1),\dots, (a_n,p_n,t_n)]=\sum_{i=1}^n f(a_i,p_i,t_i),
\end{equation}  
where  $ f :A\times [0,1]\times \Bbb{R}_+\rightarrow \mathbb{R}_+$ 
is continuous with respect to its second and third variables and for each $d=[d_1,\dots,d_n]=[(a_1,p_1,t_1),\dots, (a_n,p_n,t_n)] \in D$, 
$$d \sim [(a_\ast,1, f(a_i,p_i, t_i))_{i\in N}].$$
where, for each $(a_i,p_i,t_i) \in A\times [0,1]\times \Bbb{R}_+$, $0 \le f(a_i,p_i,t_i) \le t_i$,  $f(a_i,p_i,t_i) \le f(a_i,1,t_i)$, and $f(a_i,p_i,t_i) \le f(a_\ast,p_i,t_i)$.
\medskip 

The evaluation function (\ref{eqlinHPYE}) is obtained via an intuitive process. Each individual triple $d_i=(a_i, p_i, t_i)$ is first associated with another triple $(a_\ast,1, f(a_i,p_i, t_i))$ in which the individual enjoys full health and maximal productivity, but a lower lifetime. This converts the multidimensional evaluation problem into a unidimensional one in which the social planner only needs to focus on lifetimes (which are simply aggregated afterwards). The precise lifetime amount, $f(a_i,p_i, t_i)$, is actually a function of the original triple and it is obtained to guarantee that the social planner is indifferent between the original distribution and the new one. This is a well-defined function, due to the COMMON axioms.  

Note that this family (\ref{eqlinHPYE}) includes the previous one (\ref{tw-HPYE}). To see that, note that for each $d=[d_1,\dots,d_n]=[(a_1,p_1,t_1),\dots, (a_n,p_n,t_n)] \in D$, 
$$
E^w[d_1,\dots,d_n]= E^w[(a_\ast,1,w(a_1,p_1)t_1),\dots, (a_\ast,1,w(a_n,p_n)t_n)].
$$
That is, 
$$d \sim [(a_\ast,1, w(a_i,p_i)t_i)_{i\in N}].$$

As the next result states, the general family of evaluation functions (\ref{eqlinHPYE}) is characterized when combining the COMMON set of axioms and \textit{time invariance at full health and productivity}.

\begin{theorem}\label{thm:linHPYE}
    The following statements are equivalent:
    \begin{enumerate}
     \item $\succsim $ is represented by a HPYE evaluation function (\ref{eqlinHPYE}).
    \item $\succsim $ satisfies COMMON and TIFHP. 
    \end{enumerate}
\end{theorem}

Theorem \ref{thm:linHPYE} implies that if the social planner endorses the view that if an intervention leads to extra life years for individuals with full health and maximal productivity, it does not matter which particular individual  receives these extra life years (axiom TIFHP), then the evaluation will be via the unweighted aggregation of the HPYEs the intervention yields (which are well defined due to COMMON).
\medskip

Finally, we can define the so-called \textit{generalized HPYE} evaluation function by the unweighted aggregation of the image of HPYEs the distribution yields to a certain function.  
Formally,
\begin{equation}\label{eqgenHPYE}
 E^g[d_1,\dots,d_n]= E[(a_1,p_1,t_1),\dots, (a_n,p_n,t_n)]=\sum_{i=1}^n g(f(a_i,p_i,t_i)),
\end{equation}
where $g: \mathbb{R}_{+} \rightarrow \mathbb{R}$ is a strictly increasing and continuous function, and $ f :A\times [0,1]\times \Bbb{R}_+\rightarrow \mathbb{R}_+$ is continuous with respect to its second and third variables and for each $d=[d_1,\dots,d_n]=[(a_1,p_1,t_1),\dots, (a_n,p_n,t_n)] \in D$, 
$$d \sim [(a_\ast,1, f(a_i,p_i, t_i))_{i\in N}].$$
where, for each $(a_i,p_i,t_i) \in A\times [0,1]\times \Bbb{R}_+$, $0 \le f(a_i,p_i,t_i) \le t_i$,  $f(a_i,p_i,t_i) \le f(a_i,1,t_i)$, and $f(a_i,p_i,t_i) \le f(a_\ast,p_i,t_i)$.
\medskip

Our last result states that the generalized HPYE evaluation function is precisely characterized by the set of COMMON axioms. 
 
\begin{theorem}\label{thm:genHPYE}
    The following statements are equivalent:
    \begin{enumerate}
     \item $\succsim $ is represented by a generalized HPYE evaluation function (\ref{eqgenHPYE}).
    \item $\succsim $ satisfies COMMON.
    \end{enumerate} 
\end{theorem}

Theorem \ref{thm:genHPYE} implies that if the social planner dismisses the TIFHP axiom, while still endorsing COMMON, then the evaluation will be via a general (but strictly increasing and continuous) function of the HPYEs the intervention yields (which are well defined due to COMMON). Thus, the evaluation will not necessarily be via a simple unweighted aggregation of HPYEs, which was the consequence of the TIFHP axiom, as formalized by the evaluation function (\ref{eqlinHPYE}).

\section{Discussion} \label{sec:discussion}

The results presented above in the form of evaluation functions and their required axioms may be utilised in empirical applications, for example as part of an economic evaluation of a health care or working environment intervention. The data collection of the economic evaluation will typically involve capturing information on all individuals in the intervention and control group regarding their costs, health status and productivity level during the follow-up period of the study.

As illustrated throughout the text with the two distributions from Example 1, the choice of evaluation function matters to a large extent when it comes to rank different distributions. Instead of making that choice directly based on their functional forms, we rather believe the choice should be guided by the axioms they satisfy. Hence the interest of our axiomatic approach.

If the analyst (working on behalf of a social planner) is of the view that health effects and productivity effects are both important outcomes when assessing the benefit of the working environment intervention, evaluation functions (5)-(10) may be used for calculating the combined effects on health and productivity in the intervention and control group. The analyst may choose a specific evaluation function based on an examination of the individual sets of axioms in terms of their expected acceptability to the society where the working environment intervention will be introduced. For example, the time independence for unproductive individuals (TIUP) axiom may be considered unacceptable, as it conveys that gains in lifetime among individuals outside the labour market have no value. Rejection of the TIUP axiom would exclude the PQALY evaluation function (5). Assume further that the analyst deems the axioms time invariance at common health and productivity (TICHP) and productivity invariance at common time (PICT) to be a good reflection of the values held by society. This would identify the QALY-PALY evaluation function (7) as the appropriate function for calculating the combined effects on health and productivity in the intervention and control group. The analyst may validate the choice of the QALY-PALY evaluation function by designing hypothetical tests that reveal if a respondent agrees to the values expressed in axioms TICHP and PICT. Participants in the economic evaluation of the intervention (or a representative sample of the population) may be exposed to these hypothetical tests.\footnote{If people largely disagree with this axiom (i.e., a majority of participants in the study systematically and significantly favor individuals with a certain type of health states over individuals with another type) it shows the need for using a more flexible evaluation function (for example the QALY-PQALY). If people largely agree with this axiom (i.e., a majority is largely indifferent between adding productivity to one type or another, or roughly participants are split into those that favor persons of one type and those that favor persons of the other type) it provides support for using the QALY-PALY.}  This will be similar in nature to the numerous tests performed of the social value of health improvements (see, for example, Gyrd-Hansen 2004, Dolan et al., 2005, Lancsar et al., 2011, Robson et al., 2024).

If the QALY-PALY evaluation function is chosen, the next step is to estimate the corresponding parameters, including the one addressing the relative importance of improvements in health and productivity ($\sigma$), the quality weights of health states ($q(a_i)$), productivity levels ($p_i$) and durations ($t_i$). The latter two parameters will typically be available directly from the data collection for the economic evaluation. Quality weights of health states may be elicited from representative agents using a person trade-off method.\footnote{The reader is referred to Nord (1995), Murray et al., (1997), or \O sterdal (2009), for details about the person trade-off technique and discussion. We also acknowledge here that the descriptive validity of this technique is highly questionable (e.g., Doctor et al., 2009).} For example, each respondent would be asked which of the following 
interventions are most desirable for society:

Intervention A: $1000$ individuals obtaining $1$ year in full health and having zero productivity. 

Intervention B: $x$ individuals obtaining $1$ year in health state $a$ and having zero productivity. 

In particular, each respondent would be asked to identify the number of individuals $x$ in Intervention B to be indifferent between 
both interventions. The quality weight can then be derived as $q(a)=\frac{1000}{x}$.

The parameter $\sigma$ may be elicited using a different version of the person trade-off technique. To wit, respondents would now be asked to state which of the following interventions are most desirable for society:

Intervention C: one individual obtaining $1$ year in full health and having zero productivity. 

Intervention D: one individual obtaining $y$ years in health state $a$ and having maximum productivity.

Each respondent would then be asked to identify the duration $y$ in Intervention D to be indifferent between the two interventions. Once $y$ is identified, it follows from evaluation function (7) that 
$\sigma = \sigma q(a)y + (1- \sigma)y$, which implies $\sigma =\frac{y}{1 - q(a)y + y}$,
from which the parameter $\sigma$ can be estimated. When all parameters have been estimated, the total effects in the intervention and control group can be calculated using the PALY-QALY evaluation function (7).

\section{Concluding remarks}\label{sec:final remarks}
Some of our models, particularly the more general functional forms (9) and (10), embrace a wide spectrum of potential ethical and social concerns. But the more structured evaluation functions (1)-(8) involve a time-linear component. Thus, for each evaluation function of the form (1)-(8), if a person doubles her time (for a fixed productivity level and health state), her contribution to social welfare doubles. This standard feature is widely acknowledged in QALY and PALY studies; indeed it reflects the basic idea of proportionally ``adjusting" the years obtained based on available information about the years lived. In the context of QALY studies, more general functional forms have been proposed. In particular, an approach where the individual QALY component is transformed by a power function before being aggregated across individuals was proposed by Wagstaff (1991) and Williams (1997), among others, and axiomatically characterized by \O sterdal (2005) and Hougaard et al. (2013) in their model focusing on population distributions of life years and health states.\footnote{The related idea of performing power transformations of individual utility functions for social welfare evaluation was pioneered by Bergson (e.g., Burk 1936).} 
In a health setting, \textit{power QALY} functional forms are typically associated to a formalization of the so-called ``fair innings", a popular argument within the public health literature (e.g., Adler et al., 2021). They can also be seen as \textit{prioritarian} evaluation functions (also known as prioritarian social welfare functionals), which rank well-being vectors according to the sum of a strictly increasing and strictly concave transformation of individual well-being (e.g., Adler, 2012, 2019; Morton, 2014). Recently, Da Costa et al. (2024) have introduced a measure of population health (dubbed \textit{equivalent health-adjusted lifespan}) that is sensitive to inequality in both age-specific health and health-adjusted lifespan. It is a life years metric that nests health-adjusted life expectancy. 


In the present framework considering health states, productivity and life years, we could also impose a power transformation of the individual components in each of the structured evaluation functions (1)-(8). This would entail determining one additional parameter for the model (or perform robustness analyses for a reasonable range of possible power transformations). We could also consider some specific prioritarian evaluation functions within the general evaluation functions (9)-(10). The consequence would be to give more weight to the initial years from the point of view of a social planner. This would change the ethical and social implications of the evaluation function used. In particular, it would mean that some of the axioms would be violated, and thus we would need new (weaker) axioms to characterize such modified and more flexible functional form. An investigation into the axiomatic underpinnings of employing a power transformation (or other parametric concave transformations) in a framework incorporating individual health, productivity, and life years is left for future research.\footnote{A natural way to start would be by modifying the TICHP axiom, so as to prefer to give the extra life years to the shorter-lived individual.} 

As we mentioned in the introduction, 
there is a pressing need to protect the health and productivity of the economically active population. 
Treatments for the elderly (retired people) have no effects on (labour market) productivity. This might render QALYs more appropriate for the evaluation of these treatments. But using the same evaluation function for younger patients misses productivity effects. 
This is a possible motivation for age weights (another motivation is the fair innings argument mentioned above). Our hybrid evaluation functions, such as PQALYs, QALYs-PALYs, QALYs-PQALYs or the more general functional form (8), offer an alternative way of dealing with productivity differences, selecting the appropriate parameters therein. 

We conclude mentioning that our framework allows for alternative plausible interpretations, as well as for further generalizations. 

Regarding the former, we focused on chronic health states, for ease of exposition, but our analysis also allows to consider time-varying health. To wit, we made no assumptions regarding the domain of health states $A$ (except for the existence of a \textit{maximal} element). In particular, this allows for time trajectories rather than fixed levels of health (with the trajectory determined by $t_{i}$). 
That is, $a_i=a_i (\cdot)$, where $a_i(s)$ denotes the health status of individual $i$ at time $s\le t_{i}$. This would require a reinterpretation of some of the axioms we considered above.\footnote{ Bossert and D'Ambrosio (2013) is a nice example of an axiomatic analysis of time trajectories (streams) of wealth (rather than health).}
 
Likewise, instead of assuming that $p_i\in[0,1]$ captures the productivity of individual $i$, we could assume that it captures the probability to succeed in life that individual $i$ has. Mariotti and Veneziani (2018) refer to this as ``chances of success" and characterize a multiplicative form to evaluate social profiles of ``chances of success" (also known as ``boxes of life").\footnote{See also Mariotti and Veneziani (2012) and Alcantud et al., (2022) for characterizations of alternative functional forms in the same model.} We could also derive their characterization results in our model upon endorsing first the axiom of \textit{health independence} 
and a counterpart axiom of \textit{time independence} (not considered in this paper). We, nevertheless, acknowledge that their main axiom is one formalizing a non-interference principle that we do not consider in this paper.
\footnote{If one considers this interpretation of our model in terms of probabilities, then the problem becomes closer to one of assessing risky situations (e.g., Fleurbaey, 2010; Eden, 2020). 
} 

As for further generalizations, we stress that the scope of our theory can be enlarged to account for more general evaluations. To wit, our theory deals with the evaluation of population distributions where individuals can be characterized by two instantaneous attributes (one qualitative and one quantitative) and a duration. These attributes can indeed be interpreted as health (qualitative) and productivity (quantitative), as we do in this paper. But there are other potential interpretations (such as happiness or well-being, to name a few). 
Our results could therefore provide interesting lessons for those settings too. 

\appendix
\section*{Appendix}\label{app}
We gather in this appendix all the proofs of the results stated above. We start with the most general result, in which we characterize all the evaluation functions satisfying the COMMON axioms. We shall then be proving the remaining results adding extra axioms to COMMON. 
\subsection*{Proof of Theorem \ref{thm:genHPYE}}
Suppose first that $\succsim $ is represented by a PHEF satisfying $\eqref{eqgenHPYE}$. It is straightforward to show that ANON and SEP hold. CONT holds because $f$ and $g$ are continuous functions themselves. As $0\le f(a_i,p_i,t_i)\le t_i$, it follows that $f(a_i,p_i,0)=0\le f(a_i,p_i,t_i)$, implying both ZERO and PLD. If $t_i>t_i'$, then $f(a_{\ast},1,t_i)=t_i>t_i'=f(a_{\ast},1,t_i')$. As $d \sim [(a_{\ast},1,f(a_i,p_i,t_i))_{i\in N}]$, and $g$ is strictly increasing, it follows that $[(a_{\ast},p_i,t_i), d_{N\setminus \{i\}}]\succ [(a_{\ast},p_i,t_i'), d_{N\setminus \{i\}}]$, so LMFHP holds. Finally, as $f(a_i,p_i,t_i)\le t_i$, $f(a_i,p_i,t_i) \le f(a_i,1,t_i)$, and $f(a_i,p_i,t_i) \le f(a_\ast,p_i,t_i)$, it follows from LMFHP that $[(a_\ast,p_i, t_i),d_{N\setminus \{i\}}] \succsim d$ and $[(a_i,1, t_i),d_{N\setminus \{i\}}] \succsim d$. 
Thus, FHPS holds.

Conversely, assume now that preferences satisfy all the axioms in COMMON.  We start by showing that there exists a function 
$f: A\times [0, 1] \times \Bbb{R}_+ \rightarrow \mathbb{R}$ such that $f$ is continuous and non-decreasing with respect to its second and third variable and such that for each $d=[d_1,\dots,d_n]=[(a_1,p_1,t_1),\dots, (a_n,p_n,t_n)] \in D$, 
$$d \sim  [(a_\ast,1,f(a_i,p_i,t_i))_{i\in N}],$$ 
where $0 \le f(a_i,p_i,t_i) \le t_i$ for each $(a_i,p_i,t_i) \in A \times [0, 1] \times \Bbb{R}_+$. Note that this part of the proof follows along the lines of proofs of existence of individual preference HYEs in \O sterdal (2005) and social preference HYEs in Hougaard et al. (2013) and Moreno-Ternero et al. (2023), with and without reference lifetime, respectively.

First, we prove that for each $d\in D$ and each $i\in N$, there exists $t_i^\ast\in \Bbb{R}_+$, such that 
$$d \sim [(a_\ast,1,t_i^\ast), d_{N\setminus \{i\}}].$$

If $t_i=0$, then it follows from ZERO that $t_i^\ast=t_i=0$. Therefore, assume $t_i>0$. We prove that $t_i^\ast$ exists by contradiction. Therefore, assume that $t_i^\ast$ does not exist. Then, $T=A \cup B$, where 
$$A=\{s\in \Bbb{R}_+:d \succ [(a_\ast,1,s), d_{N\setminus \{i\}}]\}$$
$$B=\{s\in \Bbb{R}_+:[(a_\ast,1,s), d_{N\setminus \{i\}}] \succ d\}. $$

By FHPS, $[(a_\ast,1,t_i), d_{N\setminus \{i\}}] \succsim d$,  implying that either $t_i^\ast=t_i$ (a contradiction), or $t_i \in B$. Assume the latter. Thus, $B$ is a non-empty set.

By PLD and ZERO, it follows that either $t_i^\ast=0$ (a contradiction), or $0 \in A$. Again, assume the latter. Thus, $A$ is a non-empty set.

By CONT, $A$ and $B$ are open sets relative to $T$. Altogether, it follows that $T$ is not a connected set, which is a contradiction. 

Thus, $t_i^\ast$ exists, and due to LMFHP, it is uniquely determined. Finally, by SEP, we can determine each $t_i^\ast$ separately. Therefore, let $f_i:A \times [0, 1] \times \Bbb{R}_+\rightarrow \mathbb{R}$ be such that $f_i(a_i,p_i,t_i)=t_i^\ast$ for each $i\in N$. By ANON, $f_i()=f_j()=f()$ for each $i,j\in N$. 
By CONT, $f$ is continuous with respect to its second and third variable and, by the above, we know that $0 \le f(a_i,p_i,t_i) \le t_i$, so the range of $f$ is a connected subset of $\mathbb{R}$. Also, by FHPS, $f(a_i,p_i,t_i) \le f(a_i,1,t_i)$, and $f(a_i,p_i,t_i) \le f(a_\ast,p_i,t_i)$ for each $(a_i,p_i,t_i) \in A\times [0,1]\times \Bbb{R}_+$. 
Thus, 
$$d \sim  [(a_\ast,1,f(a_i,p_i,t_i))_{i\in N}],$$ 
which implies that social preferences only depend on the profile of HPYEs, and, by CONT, they do so continuously. As in the models of \O sterdal (2005), Hougaard et al. (2013), and Moreno-Ternero et al. (2023), it then follows by application of Theorem 3 in Debreu (1960) that 
$$d\succsim d' \Leftrightarrow \sum_{i=1}^n g(f(a_i,p_i,t_i))\ge \sum_{i=1}^n g(f(a_i',p_i',t_i')),$$
where $g: \mathbb{R_+} \rightarrow \mathbb{R}$ is strictly increasing.  \endproof

\subsection*{Proof of Theorem \ref{thm:linHPYE}}

Suppose first that $\succsim $ is represented by a PHEF satisfying $\eqref{eqlinHPYE}$. By Theorem \ref{thm:genHPYE}, COMMON holds. 
As for TIFHP, let $d\in D$ and $i,j\in N$ be such that $a_i=a_j=a_\ast$ and $p_i=p_j=1$. Then, for each $c>0$, 
$$[(a_\ast,1, t_i+c), d_{ N\setminus \{i\}}] =f(a_\ast,1, t_i+c)+f(a_\ast,1, t_j)+\sum_{k\in N\setminus \{i,j\}} f(a_k,p_k,t_k )=t_i+c+t_j+\sum_{k\in N\setminus \{i,j\}} f(a_k,p_k,t_k ),$$
and 
$$
[(a_\ast,1,t_j+c), d_{ N\setminus \{j\}}]=f(a_\ast,1, t_j+c)+f(a_\ast,1, t_i)+\sum_{k\in N\setminus \{i,j\}} f(a_k,p_k,t_k )=t_j+c+t_i+\sum_{k\in N\setminus \{i,j\}} f(a_k,p_k,t_k ).
$$
Thus, 
$$[(a_\ast,1, t_i+c), d_{ N\setminus \{i\}}]\sim [(a_\ast,1,t_j+c), d_{ N\setminus \{j\}}].$$

Conversely, assume now that preferences satisfy all the axioms in COMMON as well as TIFHP. Then, by Theorem \ref{thm:genHPYE}, for each pair $d,d'\in D$, 
$$d\succsim d' \Leftrightarrow \sum_{i=1}^n g(f(a_i,p_i,t_i))\ge \sum_{i=1}^n g(f(a_i',p_i',t_i')),$$
where $g: \mathbb{R_+} \rightarrow \mathbb{R}$ is strictly increasing. Now, for each pair $t_i,t_j\in \Bbb{R}_+$, and for each $c>0$, it follows by TIFHP that 
$$
g(f(a_\ast,1, t_i+c))+g(f(a_\ast,1, t_j))=g(f(a_\ast,1, t_i))+g(f(a_\ast,1, t_j+c)).
$$
Or, equivalently, 
$$
g(t_i+c)+g(t_j)=g(t_i)+g(t_j+c),
$$
In particular, 
$$
g\left(\frac{x+y}{2}\right)=\frac{g(x)+g(y)}{2},
$$
for each $x,y\ge 0$. 
As $g$ is continuous and strictly increasing, it follows from Theorem 1 in Aczel (2006, p. 43) that there exist $\alpha$ and $\beta$ such that $g(x)=\alpha x+\beta$, for each $x>0$. By LMFHP, $\alpha>0$. Consequently,  $\succsim $ is indeed represented by an evaluation function satisfying $(\ref{eqlinHPYE})$, as desired.\endproof

\subsection*{Proof of Theorem \ref{thm:tw-HPYE}}
Suppose first that $\succsim $ is represented by a PHEF satisfying $\eqref{tw-HPYE}$. As this is a special case of \eqref{eqgenHPYE}, it follows from Theorem \ref{thm:genHPYE} that COMMON holds. 
As for TICHP, let $d\in D$ and $i,j\in N$ be such that $a_i=a_j=a$ and $p_i=p_j=p$. Then, for each $c>0$, 
$$E[(a,p,t_i+c),(a,p,t_j), d_{ N\setminus \{i,j\}}] =w(a,p)(t_i+c)+w(a,p)t_j+\sum_{k\in N\setminus \{i,j\}} w(a_k,p_k)t_k,$$
and 
$$
E[(a,p,t_i),(a,p,t_j+c), d_{ N\setminus \{i,j\}}]=w(a,p)(t_j+c)+w(a,p)t_i+\sum_{k\in N\setminus \{i,j\}} w(a_k,p_k)t_k.
$$
Thus, 
$$[(a,p,t_i+c),(a,p,t_j), d_{ N\setminus \{i,j\}}]\sim [(a,p,t_i),(a,p,t_j+c), d_{ N\setminus \{i,j\}}].$$
Conversely, assume now that preferences satisfy all the axioms in the statement of Theorem \ref{thm:tw-HPYE}. Then, by Theorem \ref{thm:genHPYE}, for each pair $d,d'\in D$, 
$$d\succsim d' \Leftrightarrow \sum_{i=1}^n g(f(a_i,p_i,t_i))\ge \sum_{i=1}^n g(f(a_i',p_i',t_i')),$$
where $g: \mathbb{R_+} \rightarrow \mathbb{R}$ is strictly increasing. 
Let $\varphi:A\times [0,1]\times \Bbb{R}_+ \rightarrow \mathbb{R}$ be such that $\varphi(a_i,p_i,t_i)=g(f(a_i,p_i,t_i))$, for each $(a_i,p_i,t_i)\in A\times [0,1]\times \Bbb{R}_+$. 
Assume, without loss of generality, that $\varphi(\overline{a},\overline{p},0)=0$ for some $(\overline{a},\overline{p})\in A\times [0,1]$. 
Let $(a,p)\in A\times [0,1]$. By iterated application of TICHP and the transitivity of $\succsim$, as well as ZERO, 

\begin{align}\label{eqSum2}
\begin{split}
\sum_{i=1}^n \varphi(a,p,t_i) 
\overset{\text{[TICHP]}}= &\varphi(a,p,\sum_{i=1}^n t_i)+ (n-1)\varphi(a,p,0) \\
\overset{\text{[ZERO]}}= &\varphi(a,p,\sum_{i=1}^n t_i)+ (n-1)\varphi(\overline{a},\overline{p},0)\\
&= \varphi(a,p,\sum_{i=1}^n t_i).
\end{split}
\end{align}

In particular, $\varphi(a,p,t_1+t_2)=\varphi(a,p,t_1)+\varphi(a,p,t_2)$ for each pair $t_1,t_2\in \Bbb{R}_+$, which is precisely one of Cauchy's canonical functional equations. As $\varphi(a,p,\cdot)$ is a continuous function, it follows that the unique solutions to such an equation are the linear functions (e.g., Aczel, 2006; page 34). More
precisely, there exists a function $\hat{w}:A\times [0,1] \rightarrow \mathbb{R}$ such that
\begin{equation*}
\varphi(a,p,t)=\hat{w}(a,p)t,
\end{equation*}%
for each $(a,p)\in A\times [0,1]$, and each $t\in \Bbb{R}_+$. By PLD, $\hat{w}(a,p)\ge 0$, for each $(a,p)\in A\times [0,1]$. By LMFHP, $\hat{w}(a_{\ast},1)>0$. By CONT, $w$ is a continuous function in its second variable. It also follows from FHPS that $\hat{w}(a_{\ast},p) \ge \hat{w}(a,p)$ and $\hat{w}(a,1) \ge \hat{w}(a,p)$ for each $(a,p)\in A\times [0,1]$. To conclude, let $w:A\times [0,1] \rightarrow \mathbb{R}$ be such that $w(a,p)=\frac{\hat{w}(a,p)}{\hat{w}(a_{\ast},1)}$, for each $(a,p)\in A\times[0,1]$. Thus, it follows that $1=w(a_{\ast},1)\ge w(a,p)\ge 0$, for each $(a,p)\in A\times[0,1]$. Then, we may write: 
$$\varphi(a,p,t_i)=w(a,p)t_i,$$
where $0\le w(a,p)\le w(a_{\ast},p)\le w(a_{\ast},1)=1$, and $0\le w(a,p)\le w(a,1)\le w(a_{\ast},1)=1$ for each $(a,p)\in A\times[0,1]$, as desired. \endproof

\subsection*{Proof of Theorem \ref{QALY}}
Suppose first that $\succsim $ is represented by a PHEF satisfying $\eqref{eqQALY}$. As this is a special case of \eqref{eqgenHPYE}, it follows from Theorem \ref{thm:genHPYE} that COMMON holds. It is straightforward to see that it also satisfies PI. 
As for TICHFP, let $d\in D$ and $i,j\in N$ be such that $a_i=a_j=a$ and $p_i=p_j=1$. Then, for each $c>0$, 
$$E[(a,1,t_i+c),(a,1,t_j), d_{ N\setminus \{i,j\}}] =q(a)(t_i+c)+q(a)t_j+\sum_{k\in N\setminus \{i,j\}} q(a_k)t_k,$$
and 
$$
E[(a,1,t_i),(a,1,t_j+c), d_{ N\setminus \{i,j\}}]=q(a)(t_j+c)+q(a)t_i+\sum_{k\in N\setminus \{i,j\}} q(a_k)t_k.
$$
Thus, 
$$[(a,1,t_i+c),(a,1,t_j), d_{ N\setminus \{i,j\}}]\sim [(a,1,t_i),(a,1,t_j+c), d_{ N\setminus \{i,j\}}].$$

Conversely, assume now that preferences satisfy all the axioms in the statement of Theorem \ref{QALY}. Then, by Theorem \ref{thm:genHPYE}, for each pair $d,d'\in D$, 
$$d\succsim d' \Leftrightarrow \sum_{i=1}^n g(f(a_i,p_i,t_i))\ge \sum_{i=1}^n g(f(a_i',p_i',t_i')),$$
where $g: \mathbb{R_+} \rightarrow \mathbb{R}$ is strictly increasing. PI and TICHFP together imply TICHP. Thus, preferences satisfy all the axioms in the statement of Theorem $\ref{tw-HPYE}$. Thus, for each $d\in D$, 
$$
E[d]=E[(a_1,p_1,t_1),\dots, (a_n,p_n,t_n)]=\sum_{i=1}^n w(a_i,p_i)t_i,
$$
where $0\le w(a,p)\le w(a_{\ast},p)\le w(a_{\ast},1)=1$, and $0\le w(a,p)\le w(a,1)\le w(a_{\ast},1)=1$ for each $(a,p)\in A\times[0,1]$. Now, by PI, $w(a,p)=w(a,1)$, for each $a\in A$. Let $q:A \rightarrow \mathbb{R}$ be such that $q(a_i)=w(a_i,1)$, for each $a_i\in A$. Then, it follows that $1=q(a_{\ast})\ge q(a)\ge 0$, for each $a\in A$. And we may write: 
$$
E[d]=E[(a_1,p_1,t_1),\dots, (a_n,p_n,t_n)]=\sum_{i=1}^n q(a)t_i,
$$
where $0\le q(a)\le q(a_{\ast})=1$, for each $a\in A$, as desired. \endproof

\subsection*{Proof of Theorem \ref{gPALY}}
Suppose first that $\succsim $ is represented by a PHEF satisfying $\eqref{eqgPALY}$. As this is a special case of \eqref{eqgenHPYE}, it follows from Theorem \ref{thm:genHPYE} that COMMON holds. It is straightforward to see that it also satisfies HI. 
As for TIFHCP, let $d\in D$ and $i,j\in N$ be such that $a_i=a_j=a_{\ast}$ and $p_i=p_j=p$. Then, for each $c>0$, 
$$E[(a_{\ast},1,t_i+c),(a_{\ast},1,t_j), d_{ N\setminus \{i,j\}}] =v(1)(t_i+c)+v(1)t_j+\sum_{k\in N\setminus \{i,j\}} v(p_k)t_k,$$
and 
$$
E[(a_{\ast},1,t_i),(a_{\ast},1,t_j+c), d_{ N\setminus \{i,j\}}]=v(1)(t_j+c)+v(1)t_i+\sum_{k\in N\setminus \{i,j\}} v(p_k)t_k.
$$
Thus, as $v(1)=1$,
$$[(a_{\ast},1,t_i+c),(a_{\ast},1,t_j), d_{ N\setminus \{i,j\}}]\sim [(a_{\ast},1,t_i),(a_{\ast},1,t_j+c), d_{ N\setminus \{i,j\}}].$$
Conversely, assume now that preferences satisfy all the axioms in the statement of Theorem \ref{gPALY}. HI and TIFHCP together imply TICHP. Thus, preferences satisfy all the axioms in the statement of Theorem $\ref{tw-HPYE}$. Then, for each $d\in D$, 
$$
E[d]=E[(a_1,p_1,t_1),\dots, (a_n,p_n,t_n)]=\sum_{i=1}^n w(a_i,p_i)t_i,
$$
where $0\le w(a,p)\le w(a_{\ast},p)\le w(a_{\ast},1)=1$, and $0\le w(a,p)\le w(a,1)\le w(a_{\ast},1)=1$ for each $(a,p)\in A\times[0,1]$. Now, by HI, $w(a,p)=w(a_{\ast},p)$, for each $p\in [0,1]$. Let $v:[0,1] \rightarrow \mathbb{R}$ be such that $v(p_i)=w(a_{\ast},p_i)$, for each $p_i\in [0,1]$. Then, it follows that $1=v(1)\ge v(p)\ge 0$, for each $p\in [0,1]$. As $w$ is continuous on its second variable, $v$ is continuous too. And we may write: 
$$E[d]=E[(a_1,p_1,t_1),\dots, (a_n,p_n,t_n)]=\sum_{i=1}^n v(p_i)t_i,$$
where $0\le v(p)\le v(1)=1$, for each $p\in [0,1]$, as desired. \endproof

\subsection*{Proof of Theorem \ref{a-PALY}}

Suppose first that $\succsim $ is represented by a PHEF satisfying $\eqref{a-PALY}$. As this is a special case of \eqref{gPALY}, it follows from Theorem \ref{gPALY} that COMMON, HI, and TIFHCP hold. 
As for PIFHCT, let $d\in D$, and $i,j\in N$ with $a_i=a_j=a_{\ast}$ and $t_i=t_j=t$. Then, for each $c>0$ such that $p_i+c,p_j+c \leq 1$,

$$E[(a_\ast,p_i+c,t),(a_\ast,p_j,t), d_{ N\setminus \{i,j\}}] = (\alpha(p_i+c)+1-\alpha) t+(\alpha p_j+1-\alpha)  t+\sum_{k\in N\setminus \{i,j\}} (\alpha p_k+1-\alpha)  t_k,$$
and
$$E[(a_\ast,p_i,t),(a_\ast,p_j+c,t), d_{ N\setminus \{i,j\}}]=(\alpha p_i+1-\alpha)  t+(\alpha(p_j+c)+1-\alpha) t+\sum_{k\in N\setminus \{i,j\}} (\alpha p_k+1-\alpha)  t_k.$$
Thus, 
$$[(a_\ast,p_i+c,t),(a_\ast,p_j,t), d_{ N\setminus \{i,j\}}]\sim [(a_\ast,p_i,t),(a_\ast,p_j+c,t), d_{ N\setminus \{i,j\}}].$$

Conversely, assume now that preferences satisfy all the axioms in the statement of Theorem \ref{affPALY}. Then, it satisfies all the axioms in Theorem \ref{gPALY}. Thus, for each pair $d,d'\in D$, 
$$d\succsim d' \Leftrightarrow \sum_{i=1}^n v(p_i)t_i\ge \sum_{i=1}^n v(p_i')t_i',$$
where $v: [0,1] \rightarrow \mathbb{R}$ is such that $0\le v(p)\le v(1)=1$, for each $p\in [0,1]$. 

Now, by PIFHCT, it follows that $v(p_i+c)-v(p_i)=v(p_j+c)-v(p_j)$, for each pair $p_i,p_j\in [0,1]$ and each $c>0$ such that $p_i+c,p_j+c\in [0,1]$. In particular, 
$$
v\left(\frac{x+y}{2}\right)=\frac{v(x)+v(y)}{2},
$$
for each $x,y\in [0,1]$. 
As $v$ is continuous and bounded, it follows from Theorem 1 in Aczel (2006, p. 43) that there exist $\alpha,\beta\in \mathbb{R}$ such that $v(x)=\alpha x+\beta$, for each $x\in [0,1]$. 
Also, $0\le \alpha p+\beta\le \alpha+\beta=1$, for each $p\in [0,1]$. Thus, $\beta=1-\alpha$. And, as $\alpha p\le \alpha$, for each $p\in [0,1]$, it also follows that $\alpha\ge 0$. By PLD, $\beta=1-\alpha\ge 0$. Thus, $\alpha\le 1$. 
Consequently, $\succsim $ is indeed represented by an evaluation function satisfying $(\ref{a-PALY})$, as desired.\endproof

\subsection*{Proof of Theorem \ref{PALY}}

Suppose first that $\succsim $ is represented by a PHEF satisfying $\eqref{PALY}$. As this is a special case of \eqref{gPALY}, it follows from Theorem \ref{gPALY} that COMMON, HI, and TIFHCP hold. It is straightforward to see that it also satisfies TIUP. As for PIFHCT, let $d\in D$, and $i,j\in N$ with $t_i=t_j=t$. Then, for each $c>0$ such that $p_i+c,p_j+c \leq 1$,

$$E[(a_\ast,p_i+c,t),(a_\ast,p_j,t), d_{ N\setminus \{i,j\}}] = (p_i+c)t+p_j t+\sum_{k\in N\setminus \{i,j\}} p_k t_k,$$
and
$$E[(a_\ast,p_i,t),(a_\ast,p_j+c,t), d_{ N\setminus \{i,j\}}]=p_i t+p_j(t+c)+\sum_{k\in N\setminus \{i,j\}} p_k t_k.$$
Thus, 
$$[(a_\ast,p_i+c,t),(a_\ast,p_j,t), d_{ N\setminus \{i,j\}}]\sim [(a_\ast,p_i,t),(a_\ast,p_j+c,t), d_{ N\setminus \{i,j\}}].$$

Conversely, assume now that preferences satisfy all the axioms in the statement of Theorem \ref{PALY}. Then, it satisfies all the axioms in Theorem \ref{gPALY}. Thus, for each pair $d,d'\in D$, 
$$d\succsim d' \Leftrightarrow \sum_{i=1}^n v(p_i)t_i\ge \sum_{i=1}^n v(p_i')t_i',$$
where $v: [0,1] \rightarrow \mathbb{R}$ is such that $0\le v(p)\le v(1)=1$, for each $p\in [0,1]$. 

Now, by PIFHCT, it follows that $v(p_i+c)-v(p_i)=v(p_j+c)-v(p_j)$, for each pair $p_i,p_j\in [0,1]$ and each $c>0$ such that $p_i+c,p_j+c\in [0,1]$. In particular, 
$$
v\left(\frac{x+y}{2}\right)=\frac{v(x)+v(y)}{2},
$$
for each $x,y\in [0,1]$. 
As $v$ is continuous and bounded, it follows from Theorem 1 in Aczel (2006, p. 43) that there exist $\alpha,\beta\in \mathbb{R}$  such that $v(x)=\alpha x+\beta$, for each $x\in [0,1]$. By TIUP, $v(0)=0$, and thus $\beta=0$. Thus, $1=v(1)=\alpha$. Consequently, $\succsim $ is indeed represented by an evaluation function satisfying $(\ref{PALY})$, as desired.\endproof

\subsection*{Proof of Theorem \ref{thm:M-PQALY}}
Suppose first that $\succsim $ is represented by a PHEF satisfying $\eqref{eqM-PQALY}$. As this is a special case of \eqref{tw-HPYE}, it follows from Theorem \ref{thm:tw-HPYE} that COMMON and TICHP hold. It is straightforward to see that it also satisfies TIUP. 
As for PICHT, let $d\in D$, and $i,j\in N$ with $a_i=a_j=a$ and $t_i=t_j=t$. Then, for each $c>0$ such that $p_i+c,p_j+c \leq 1$,
$$E[(a,p_i+c,t),(a,p_j,t), d_{ N\setminus \{i,j\}}] = q(a)(p_i+c)t+q(a)p_j t+\sum_{k\in N\setminus \{i,j\}} q(a_k)p_k t_k,$$
and
$$E[(a,p_i,t),(a,p_j+c,t), d_{ N\setminus \{i,j\}}]=q(a)p_i t+q(a)p_j(t+c)+\sum_{k\in N\setminus \{i,j\}} q(a_k)p_k t_k.$$
Thus, 
$$[(a,p_i+c,t),(a,p_j,t), d_{ N\setminus \{i,j\}}]\sim [(a,p_i,t),(a,p_j+c,t), d_{ N\setminus \{i,j\}}].$$

Conversely, assume now that preferences satisfy all the axioms in the statement of Theorem \ref{thm:M-PQALY}. Then, they also satisfy the axioms of Theorem \ref{thm:tw-HPYE}. Thus, for each pair $d,d'\in D$, 
$$d\succsim d' \Leftrightarrow \sum_{i=1}^n w(a_i,p_i)t_i\ge \sum_{i=1}^n w(a_i',p_i')t_i',$$ 
where $ w :A\times [0,1] \rightarrow [0,1]$ is a continuous function in its second variable and such that $0\le w(a,p)\le w(a_{\ast},p)\le w(a_{\ast},1)=1$, and $0\le w(a,p)\le w(a,1)\le w(a_{\ast},1)=1$ for each $(a,p)\in A\times[0,1]$. 

For each $a\in A$, Let $w^{a}: [0,1] \rightarrow \mathbb{R}_+$ be such that $w^{a}(p)=w(a,p)$, for each $p\in [0,1]$. Then, $w^{a}$ is a continuous function and, by PICHT, such that $w^{a}(p_i+c)+w^{a}(p_j)=w^{a}(p_i)+w^{a}(p_j+c)$, for each pair $p_i,p_j\in [0,1]$ and each $c>0$ such that $p_i+c,p_j+c\in [0,1]$. In particular, 
$$
w^{a}\left(\frac{x+y}{2}\right)=\frac{w^{a}(x)+w^{a}(y)}{2},
$$
for each $x,y\in [0,1]$. 
Thus, by Theorem 1 in Aczel (2006, p. 43), 
there exist $\alpha,\beta\in \Bbb{R}$ such that $w^{a}(x)=\alpha x+\beta$, for each $x\in[0,1]$. Consequently, there exist $q,r :A\rightarrow\mathbb{R}$ such that $w(a,p)=q(a)p+r(a)$, for each $p\in [0,1]$, and each $a\in A$. By TIUP, $w(a,0)=0$ for each $a\in A$, and thus $r(a)=0$ for each $a\in A$. Consequently, $w(a,p)=q(a)p$, for each $p\in [0,1]$ and, therefore, $\succsim $ is indeed represented by an evaluation function satisfying $(\ref{eqM-PQALY})$, as desired.\endproof

\subsection*{Proof of Theorem \ref{thm:A-PQALY}}
Suppose first that $\succsim $ is represented by a PHEF satisfying $\eqref{eqA-PQALY}$. As this is a special case of \eqref{tw-HPYE}, it follows from Theorem \ref{thm:tw-HPYE} that COMMON and TICHP hold. As for PICHT, let $d\in D$, and $i,j\in N$ with $a_i=a_j=a$ and $t_i=t_j=t$. Then, for each $c>0$ such that $p_i+c,p_j+c \leq 1$,
$E[(a,p_i+c,t),(a,p_j,t), d_{ N\setminus \{i,j\}}] = \delta (q(a)t +q(a)t)+(1-\delta)(q(a)(p_i+c)t+q(a)p_j t)+ \sum_{k\in N\setminus \{i,j\}} \delta q(a_k)t_k+(1-\delta) r(a_k)p_kt_k$, 
and $E[(a,p_i,t),(a,p_j+c,t), d_{ N\setminus \{i,j\}}]=\delta (q(a)t +q(a)t)+(1-\delta)(q(a)p_i t+q(a)(p_j+c) t)+\sum_{k\in N\setminus \{i,j\}} \delta q(a_k)t_k+(1-\delta) r(a_k)p_kt_k$. 
Thus, 
$$[(a,p_i+c,t),(a,p_j,t), d_{ N\setminus \{i,j\}}]\sim [(a,p_i,t),(a,p_j+c,t), d_{ N\setminus \{i,j\}}].$$

Conversely, assume now that preferences satisfy all the axioms in the statement of Theorem \ref{thm:A-PQALY}. Then, they also satisfy the axioms in Theorem \ref{thm:tw-HPYE}. Thus, for each pair $d,d'\in D$, 
$$d\succsim d' \Leftrightarrow \sum_{i=1}^n w(a_i,p_i)t_i\ge \sum_{i=1}^n w(a_i',p_i')t_i',$$ 
where $ w :A\times [0,1] \rightarrow [0,1]$ is a continuous function in its second variable and such that $0\le w(a,p)\le w(a_{\ast},p)\le w(a_{\ast},1)=1$, and $0\le w(a,p)\le w(a,1)\le w(a_{\ast},1)=1$ for each $(a,p)\in A\times[0,1]$. 

For each $a\in A$, Let $w^{a}: [0,1] \rightarrow \mathbb{R}_+$ be such that $w^{a}(p)=w(a,p)$, for each $p\in [0,1]$. Then, $w^{a}$ is a continuous function and, by PICHT, such that $w^{a}(p_i+c)+w^{a}(p_j)=w^{a}(p_i)+w^{a}(p_j+c)$, for each pair $p_i,p_j\in [0,1]$ and each $c>0$ such that $p_i+c,p_j+c\in [0,1]$. In particular, 
$$
w^{a}\left(\frac{x+y}{2}\right)=\frac{w^{a}(x)+w^{a}(y)}{2},
$$
for each $x,y\in [0,1]$. 
Thus, by Theorem 1 in Aczel (2006, p. 43), 
there exist $\alpha',\beta'\in\mathbb{R}$ such that $w^{a}(x)=\alpha x+\beta$, for each $x\in [0,1]$. Consequently, there exist $\alpha,\beta :A\rightarrow\mathbb{R}$ such that $w(a,p)=\alpha(a)p+\beta(a)$, for each $p\in [0,1]$, and each $a\in A$. Note that $1=w(a_{\ast},1)=\alpha(a_{\ast})+\beta(a_{\ast})$. We distinguish several cases.

Case 1. $\alpha(a_{\ast})=0$. 

In this case, $\beta(a_{\ast})=1$. Furthermore, $0\le \alpha(a)p+\beta(a)\le \alpha(a_{\ast})p+\beta(a_{\ast})=1$, and $0\le \alpha(a)p+\beta(a)\le \alpha(a)+\beta(a)\le1$ for each $(a,p)\in A\times[0,1]$. Thus, $\alpha(a)\ge 0=\alpha(a_{\ast})$ for each $a\in A$. Thus, by FHPS,  $\alpha(a)= 0$ for each $a\in A$. Therefore, $w(a,p)=\beta(a)$, for each $(a,p)\in A\times[0,1]$, and $0\le \beta(a)\le \beta(a_{\ast})=1$. If we let $q:A\rightarrow [0,1]$ such that $q(a)=\beta (a)$, for each $a\in A$, it follows that $\succsim $ is represented by an evaluation function satisfying $(\ref{QALY})$. Or, equivalently, by an evaluation function satisfying $(\ref{eqA-PQALY})$, with $\delta=1$.

Case 2. $\beta(a_{\ast})=0$. 

In this case, $\alpha(a_{\ast})=1$. Furthermore, $0\le \alpha(a)p+\beta(a)\le p\le1$, and $0\le \alpha(a)p+\beta(a)\le \alpha(a)+\beta(a)\le1$ for each $(a,p)\in A\times[0,1]$. Thus, $\alpha(a)\ge 0=\alpha(a_{\ast})$ for each $a\in A$. Thus, if $p=0$, we obtain $0\le \beta(a)\le 0$, for each $a\in A$. Therefore, $w(a,p)=\alpha(a)p$, for each $(a,p)\in A\times[0,1]$, and $0\le \alpha(a)\le \alpha(a_{\ast})=1$. If we let $q:A\rightarrow [0,1]$ such that $q(a)=\alpha (a)$, for each $a\in A$, it follows that $\succsim $ is represented by an evaluation function satisfying $(\ref{eqM-PQALY})$. Or, equivalently, by an evaluation function satisfying $(\ref{eqA-PQALY})$, with $\delta=0$.

Case 3. $\alpha(a_{\ast})\ne 0\ne\beta(a_{\ast})$. 

Let $q,r:A\rightarrow \mathbb{R}$ be such that $r(a)=\frac{\alpha(a)}{\alpha(a_{\ast})}$ and $q(a)=\frac{\beta(a)}{\beta(a_{\ast})}$, for each $a\in A$. Then, by FHPS, $1=q(a_{\ast})\ge q(a)\ge 0$, and $1=r(a_{\ast})\ge r(a)\ge 0$. Furthermore, $0<\alpha(a_{\ast}),\beta(a_{\ast})<1$. Now, if we let 
$\delta=\beta(a_{\ast})\in (0,1)$, we have $w(a,p)=\delta q(a)+(1-\delta) r(a)p$, for each $p\in [0,1]$, and each $a\in A$. Thus, for each pair $d,d'\in D$, 
$$d\succsim d' \Leftrightarrow \delta\sum_{i=1}^n q(a_i)t_i+(1-\delta)\sum_{i=1}^n r(a_i)p_it_i\ge \delta\sum_{i=1}^n q(a'_i)t'_i+(1-\delta)\sum_{i=1}^n r(a'_i)p'_it'_i,
$$
where $q,r:A\rightarrow [0,1]$ are such that $1=q(a_{\ast})\ge q(a)\ge 0$, and $1=r(a_{\ast})\ge r(a)\ge 0$ and $\delta\in (0,1)$. Consequently,  $\succsim $ is indeed represented by an evaluation function satisfying $(\ref{eqA-PQALY})$, with $\delta\in (0,1)$ as desired.\endproof

\subsection*{Proof of Theorem \ref{thm:QALY-PALY}}
Suppose first that $\succsim $ is represented by a PHEF satisfying $\eqref{QALY-PALY}$. As this is a special case of \eqref{eqA-PQALY}, it follows from Theorem \ref{thm:A-PQALY} that COMMON and TICHP hold. As for PICT, let $d\in D$, and $i,j\in N$ with $t_i=t_j=t$. Then, for each $c>0$ such that $p_i+c,p_j+c \leq 1$,
$$E[(a_i,p_i+c,t),(a_j,p_j,t), d_{ N\setminus \{i,j\}}] = \sigma (q(a_i)t +q(a_j)t)+(1-\sigma)((p_i+c)t+p_j t)+\sum_{k\in N\setminus \{i,j\}} \sigma q(a_k)t_k+(1-\sigma) p_kt_k,$$
and
$$E[(a_i,p_i,t),(a_j,p_j+c,t), d_{ N\setminus \{i,j\}}]=\sigma (q(a_i)t +q(a_j)t)+(1-\sigma)(p_i t+(p_j+c) t)+\sum_{k\in N\setminus \{i,j\}} \sigma q(a_k)t_k+(1-\sigma) p_kt_k.$$
Thus, 
$$[(a_i,p_i+c,t),(a_j,p_j,t), d_{ N\setminus \{i,j\}}]\sim [(a_i,p_i,t),(a_j,p_j+c,t), d_{ N\setminus \{i,j\}}].$$

Conversely, assume now that preferences satisfy all the axioms in the statement of Theorem \ref{thm:QALY-PALY}. Then, they also satisfy the axioms in Theorem \ref{thm:A-PQALY} (note that PICT implies PICHT). Thus, for each pair $d,d'\in D$, 
$$d\succsim d' \Leftrightarrow \delta\sum_{i=1}^n q(a_i)t_i+(1-\delta)\sum_{i=1}^n r(a_i)p_it_i\ge \delta\sum_{i=1}^n q(a'_i)t'_i+(1-\delta)\sum_{i=1}^n r(a'_i)p'_it'_i,$$ 
where $q,r:A\rightarrow [0,1]$ are such that $1=q(a_{\ast})\ge q(a)\ge 0$, and $1=r(a_{\ast})\ge r(a)\ge 0$ and $\delta\in [0,1]$. 

By PICT, 
$$[(a_i,p_i+c,t),(a_j,p_j,t), d_{ N\setminus \{i,j\}}]\sim [(a_i,p_i,t),(a_j,p_j+c,t), d_{ N\setminus \{i,j\}}],$$
for each $d\in D$, and $i,j\in N$ with $t_i=t_j=t$. Now, as 
$E[(a_i,p_i+c,t),(a_j,p_j,t), d_{ N\setminus \{i,j\}}] = \delta (q(a_i)t +q(a_j)t)+(1-\delta)(r(a_i)(p_i+c)t+r(a_j)p_j t)+\sum_{k\in N\setminus \{i,j\}} \delta q(a_k)t_k+(1-\sigma) r(a_k)p_kt_k$, 
and
$E[(a_i,p_i,t),(a_j,p_j+c,t), d_{ N\setminus \{i,j\}}]=\delta (q(a_i)t +q(a_j)t)+(1-\delta)(r(a_i)p_i t+r(a_j)(p_j+c) t)+\sum_{k\in N\setminus \{i,j\}} \delta q(a_k)t_k+(1-\sigma) r(a_k)p_kt_k$. 
Thus, 
$r(a_i)=r(a_j)$, for each pair $a_i,a_j\in A$. As, $1=r(a_{\ast})$, it follows that $r(a)=1$, for each $a\in A$. 
Thus, letting $\sigma=\delta$, we obtain that $\succsim $ is indeed represented by an evaluation function satisfying $(\ref{QALY-PALY})$ as desired.\endproof

\bigskip


\end{document}